\documentclass[twocolumn,showpacs,preprintnumbers,amsmath,amssymb,prb]{revtex4}

\usepackage{graphicx}
\usepackage{dcolumn}
\usepackage{bm}
\usepackage{color}
\usepackage{epsfig}

\begin{document}

\newcommand{\commute}[2]{\left[#1,#2\right]}
\newcommand{\bra}[1]{\left\langle #1\right|}
\newcommand{\ket}[1]{\left|#1\right\rangle }
\newcommand{\anticommute}[2]{\left\{  #1,#2\right\}  }
\renewcommand{\arraystretch}{2}
\title{Nuclear spin state narrowing via gate--controlled Rabi
  oscillations in a double quantum dot} 

\author{D. Klauser}
\author{W. A. Coish}
\author{Daniel Loss}
\affiliation{Department of Physics and Astronomy, University of Basel,
Klingelbergstrasse 82, CH-4056 Basel, Switzerland}

\date{\today}

\begin{abstract}
We study spin dynamics for two electrons confined to a double quantum dot under
the influence of an oscillating exchange interaction. This leads to
driven Rabi oscillations between the $\ket{\uparrow\downarrow}$--state and the 
$\ket{\downarrow\uparrow}$--state of the two--electron system. The width of the Rabi resonance 
is proportional to the amplitude of the oscillating exchange. A measurement of the Rabi resonance
allows one to narrow the distribution of nuclear spin states and thereby to prolong the spin
decoherence time. Further, we study decoherence of the
two-electron states due 
to the hyperfine interaction and give requirements on the parameters of the system in order
to initialize in the $\ket{\uparrow\downarrow}$--state and to perform a $\sqrt{\mathrm{SWAP}}$
operation with unit fidelity. 
\end{abstract}
\maketitle

\section{Introduction}

One of the important proposals for quantum information processing in solid--state systems is
the spin--qubit proposal for quantum computing with electron spins in quantum dots 
\cite{Loss:1998a}. Much effort has been put into the realization of this proposal leading 
to exciting theoretical \cite{Engel:2005a} and experimental achievements. 
\cite{Cerletti:2005a,Ono:2002a,Ono:2004a,Petta:2005a,Hatano:2005a,Koppens:2005a,Petta:2005b} 
Still many challenges remain such as decoherence and the implementation of
single--qubit gates. 

A major obstacle to quantum computation with the quantum--dot spin qubit is
decoherence due to the coupling of the qubit to its environment. The hyperfine 
interaction between the electron spin and the nuclear spins present in all 
III-V semiconductors \cite{Schliemann:2003a} leads to the strongest 
decoherence effect 
\cite{Burkard:1999a,Erlingsson:2001a,Erlingsson:2002a,Khaetskii:2002a,Merkulov:2002a,
Khaetskii:2003a,Coish:2004a,Coish:2005a,Petta:2005b}. 
Experiments \cite{Bracker:2005a,Dutt:2005a,Koppens:2005a,Petta:2005b} have yielded values 
for the free--induction spin dephasing time $T_2^*$ that are
consistent with $T_2^*\sim \sqrt{N}/A \sim  
10 \mathrm{ns}$ \cite{Khaetskii:2002a,Merkulov:2002a,Khaetskii:2003a} for $N = 10^6$ and 
$A=90\mu \mathrm{eV}$ in GaAs, where $N$ is the number of nuclei within one quantum dot
Bohr radius and $A$ characterizes the hyperfine coupling strength.\cite{Paget:1977a}
This is to be contrasted to potential spin--echo envelope decay, which may be much larger. 
\cite{Sousa:2003a,Shenvi:2005a,Yao:2005a} With a two--qubit switching time of 
$\tau_s\sim 50$ps\cite{Burkard:1999a} this only allows $\sim 10^2$ gate operations within $T_2^*$,
which
falls short (by a factor of $10$ to $10^2$) of current requirements for efficient quantum 
error correction. \cite{Steane:2003a}

There are several ways to overcome the problem of hyperfine-induced decoherence, of 
which measurement and thus projection of the nuclear spin state seems to be the most
promising one\cite{Coish:2004a}. Other methods include polarization
\cite{Burkard:1999a,Khaetskii:2003a,Imamoglu:2003a,Coish:2004a} of the nuclear spins and
spin echo techniques.\cite{Coish:2004a,Shenvi:2005a,Petta:2005b} However, in order to 
extend the decay time  by an order of magnitude through polarization 
of the nuclear spins, a polarization of  above 99\% is required,\cite{Coish:2004a} but the best 
result so far reached is only $\sim$60\%  in quantum dots. \cite{Bracker:2005a,Cerletti:2005a} 
With spin-echo techniques, gate operations still must be performed within the single--spin 
free--induction decay time, which requires faster gate operations. A projective
measurement of the nuclear spin state leads to an extension of the
free--induction decay time for the spin. This extension is only
limited by the ability to do a strong measurement since the longitudinal nuclear spin in a 
quantum dot is expected to 
survive up to the spin diffusion time, which is on the order of seconds for nuclear spins
surrounding donors in GaAs. \cite{Paget:1982a}

The implementation of quantum computation schemes requires coherent control of the qubits. 
Rabi oscillations between the two qubit states are an important signature of coherence 
and thus observation of controlled Rabi oscillations is an important intermediate step
in the experimental implementation of quantum information processors. Despite recent 
experimental achievements, \cite{Cerletti:2005a,Petta:2005b} there has
still been no experimental 
observation of driven Rabi oscillations for a system of two
quantum--dot spin qubits. What
has been observed is electron spin resonance via g-tensor modulation in a bulk semiconductor. 
\cite{Kato:2003a}

In the quantum--dot spin qubit proposal, two--qubit gates are realized through tuning of the 
exchange coupling $J$ between the two
spins.\cite{Loss:1998a,Burkard:1999a} The splitting between
singlet and triplet states of the two--electron system is given by the
exchange coupling $J$ and in devices such as those 
in Refs. \onlinecite{Petta:2005b} and \onlinecite{Koppens:2005a}, $J$ 
can be controlled through gate voltages. Petta \emph{et al.} \cite{Petta:2005b} have
 recently managed to implement the $\sqrt{\mathrm{SWAP}}$--gate in their setup. 
However, in order to implement single--qubit gates, control over local
magnetic fields or g--factors is required.\cite{Burkard:1999a}

As we will show in Sec.\ref{sec:Rabi}, an oscillating exchange $J(t)$ induces Rabi oscillations 
between the states $\ket{\uparrow\downarrow}$ and
$\ket{\downarrow\uparrow}$ of two electron spins (one electron in each dot). The amplitude of
these oscillations is resonant on the splitting between $\ket{\uparrow\downarrow}$ and 
$\ket{\downarrow\uparrow}$ and the width of this resonance is
proportional to  the amplitude $j$
of the oscillating component of $J(t)=J_0+j\cos(\omega t)$, where
$\omega$ is the driving frequency. Since the splitting depends on the state
of the nuclear system, a measurement of the resonance is also a measurement of 
the state of the nuclear spins and thus provides a way to narrow the quantum distribution of the 
nuclear spin states. This narrowing of the spin state is one possible 
solution to suppress hyperfine--induced decoherence in quantum--dot
spin qubits \cite{Coish:2004a}.  
It has been proposed to measure the nuclear spin polarization using a phase estimation
method. \cite{Giedke:2005a} In the ideal case, phase estimation yields one bit of information
about the nuclear--spin system for each perfectly measured electron. Optical methods have
also been proposed. \cite{Stepanenko:2005a} The all--electrical method 
we present here can be applied with current technology. 

The rest of this paper is organized as follows. In Sec. \ref{sec:Rabi} we show that an
oscillating exchange leads to driven Rabi oscillations and calculate the resonance linewidth. 
In Sec. \ref{sub:squeezing} we propose a method to narrow the
distribution of the nuclear spin states.  
in Sec. \ref{sec:SzZero} we consider decoherence induced through the hyperfine 
interaction for a static exchange coupling $J$. We use these results in Sec.\ref{sec:SWAP}
to analyze under which conditions we reach unit fidelity for the
initialization to the state $\ket{\uparrow\downarrow}$ and a
$\sqrt{\mathrm{SWAP}}$ operation. \cite{Loss:1998a}
Sec. \ref{conclusion} contains a summary of our results.

\section{\label{sec:Rabi}Oscillating Exchange and ESR}

In this section we show that under suitable conditions an oscillating exchange interaction
may be used to induce Rabi oscillations in a system of two electrons confined to a double 
quantum dot like those in Refs. 
\onlinecite{Petta:2005a, Petta:2005b,Koppens:2005a,Hatano:2005a}.

We denote by $\mathbf{h}_i=(h_i^x,h_i^y,h_i^z),\,\,i=1,2$, the collective quantum nuclear spin 
operator, the ``Overhauser operator'', in dot one and two, respectively, and write 
$\delta h^z=\frac{1}{2}(h_1^z-h_2^z)$. The collective quantum nuclear spin
operator $\mathbf{h}_i$ is defined as $\mathbf{h}_i=\sum_k A^i_k\mathbf{I}_k$, where
$\mathbf{I}_k$ is the nuclear spin operator for a nucleus of total spin
$I$ at lattice site $k$, and the hyperfine coupling constants are
given by $A^i_k=v A |\psi^i_0(\mathbf{r}_k)|^2$, where $v$ is the volume
of a unit cell containing one nuclear spin, A characterizes the
hyperfine coupling strength, and $\psi^i_0(\mathbf{r}_k)$ is the
single-particle envelope wavefunction of the electron evaluated at site $k$.
Further, $\langle\mathcal{O}\rangle_{\mathrm{rms}}
=\bra{\psi_I} \mathcal{O}^2\ket{\psi_I}^{1/2}$ is the root--mean--square expectation value 
of the operator $\mathcal{O}$ with respect to the nuclear spin state $\ket{\psi_I}$.
We assume that the Zeeman splitting 
$\epsilon_z=g\mu_BB$ induced by a uniform applied magnetic field $\mathbf{B}=(0,0,B)$, $B>0$, 
is much larger than $\langle \delta \mathbf{h}\rangle_{\mathrm{rms}}$ and $\langle\mathbf{h}_i
\rangle_{\mathrm{rms}}$. Under these conditions the relevant spin Hamiltonian becomes block 
diagonal  with blocks labeled by the total electron spin projection
along the magnetic field $S^z$. 
In the subspace of $S^z=0$ the Hamiltonian can be written as ($\hbar=1$) \cite{Coish:2005a} 
\begin{equation}\label{Hamiltonian}
H_{0}=\frac{J}{2}\left(1+\tau^{z}\right)+\delta h^{z}\tau^{x}+\delta b^z \tau^x.
\end{equation}
Here, $J$ is the Heisenberg exchange coupling between electron spins on the two dots
and $\delta b^z$ the inhomogeneity of an externally applied classical
static magnetic field 
which we add in addition to the treatment in Ref. \onlinecite{Coish:2005a}. Further,
$\pmb{\tau}=(\tau^{x},\tau^{y},\tau^{z})$ is the vector of Pauli matrices in the basis of
$S^z=0$ singlet $\ket{S}$ and triplet $\ket{T_0}$  ($\ket{S}\to\ket{\tau^z=-1},
\ket{T_0}\to\ket{\tau^z=+1}$). It has been proposed to use two pseudo-spin
states such as $\ket{S}$ and $\ket{T_0}$ as a logical qubit. \cite{Levy:2002a}

We assume a time--dependent exchange of the form
\begin{equation}\label{Eq:joft}
J=J(t)=J_0+j \cos(\omega t).
\end{equation}
The operator $\delta h^z$ commutes with the Hamiltonian at all times. Thus,
if the nuclear--spin system is in an eigenstate $\ket{n}$ of $\delta h^z$ with 
$\delta h^z \ket{n}=\delta h^z_n \ket{n}$,
we have $H \ket{\psi}= H_n \ket{\psi_e}\otimes \ket{n}$, where in $H_n$ the operator
$\delta h^z$ has been replaced by $\delta h^z_n$ and $\ket{\psi_e}$ is the electron
spin part of the wave function. 
In order to bring $H_n$ to a form that is very similar to the standard ESR (electron 
spin resonance)  Hamiltonian
\cite{Slichter:1980a} 
($H_{ESR}=-\frac{1}{2}\epsilon_z\sigma_z-\frac{1}{2}\Delta_x \cos (\omega t)\sigma_x$)
we perform a unitary transformation $U_1=\exp(-i\frac{\pi}{4}\tau^y)$ which is 
just a rotation about the $y$-axis in a Bloch--sphere picture. Also introducing 
$\Omega_n=2(\delta h^z_n+\delta b^z)$, the above Hamiltonian becomes
\begin{equation}
\tilde{H}_n=U_1H_n U_1^{\dagger} = \frac{J_0}{2}\tau^x+\frac{j}{2}\cos(\omega t)\tau^x-
\frac{1}{2}\Omega_n\tau^z.
\end{equation}
The Pauli matrices are now given in the new
basis of $\ket{\downarrow\uparrow}=\ket{\tau^z=1}=\ket{+}$ and $\ket{\uparrow\downarrow}=
\ket{\tau^z=-1}=\ket{-}$. For $J_0=0$ this is just the standard ESR
Hamiltonian. We have evaluated  
pseudo--spin dynamics under this Hamiltonian in a rotating wave
approximation close to resonance       
for $j\ll\Omega_n$. 
When we treat the $J_0$--term as a perturbation and calculate the transition probability 
between unperturbed eigenstates of the Hamiltonian we find that it is proportional to 
$J_0^2/\Omega_n^2$ and we may thus neglect 
this term close to resonance and if $J_0\ll \Omega_n$. Hence, we are
left with the standard ESR Hamiltonian which leads to Rabi
oscillations. Initializing the two--electron system in the state
$\ket{\downarrow\uparrow}=\ket{+}$ (which can be done as
proposed  in Sec. \ref{sec:SWAP})
we obtain for the expectation value of $\tau^z(t)$:
\begin{eqnarray}\label{eq:tauz}
\langle\tau^z(t)\rangle_n&=& \bra{n}\otimes \bra{+}\tau^z(t)
\ket{+}\otimes \ket{n} \nonumber \\
&=&\frac{(\Omega_n-\omega)^2+(j/2)^2 \cos\left(\omega^{\prime}t\right)}
{(\Omega_n-\omega)^2+(j/2)^2},\\
& &\omega^{\prime}=2\sqrt{(\Omega_n-\omega)^2+\left(j/2\right)^2},\label{eq:omegaprime}\\
& &j\ll \Omega_n,\,\,\, J_0\ll \Omega_n,\,\,\, |\Omega_n-\omega|\ll \Omega_n.
\end{eqnarray}
For $\omega=\Omega_n$ the system undergoes coherent Rabi oscillations between the
states $\ket{+}$ and $\ket{-}$ with a frequency of $j$. 
Averaged over time, the expectation value of $\tau^z$ is
\begin{equation}\label{Eq:tauztimeaverage}
\langle\langle\tau^z\rangle_n\rangle=\lim_{T \to \infty} \frac{1}{T}\int^T_0
\langle\tau^z(t)\rangle_n dt=\frac{(\Omega_n-\omega)^2}{(\Omega_n-\omega)^2
+(j/2)^2}.
\end{equation}
In order to measure the time--averaged value $\langle\langle\tau^z\rangle_n\rangle$ 
the measurement time must
be much larger than the period of Rabi oscillations ($ \sim 1/j$ on resonance).
$1-\langle\langle\tau^z\rangle_n\rangle$ has a Lorentzian lineshape with a 
full width at half maximum (FWHM) of $j$.
Most importantly, the resonance frequency 
depends on the nuclear--spin eigenstate through $\Omega_n=2(\delta h_n^z+\delta b^z)$ and 
thus a measurement of the resonance will determine $\delta h_n^z$.

\subsection{Superposition of nuclear--spin eigenstates \label{sub:superposition}}

Before a measurement on the nuclear--spin system is performed, there is
no reason for the nuclear--spin system to be in an eigenstate of
$\delta h^z$, but it is most likely in some generic superposition of these
eigenstates. Thus, we now investigate how the resonance changes if we
consider the nuclear--spin system to be in a superposition of
eigenstates of the collective nuclear spin operator $\delta h^z$. 

At $t=0$ we fix the electron system in the state $\ket{\downarrow\uparrow}=\ket{+}$ while the
nuclear--spin system is in an arbitrary state:
$\rho(0)=\rho_e(0)\otimes \rho_I(0)$ with
\begin{eqnarray}
\rho_e(0)&=& \ket{+}\bra{+},\\
\rho_I(0)&=& \sum_ip_i\ket{\psi_I^i}\bra{\psi_I^i};\,\,\,\ket{\psi_{I}^i}=\sum_na_n^i\ket{n},
\end{eqnarray}
where the $a_n^i$ satisfy the normalization condition $\sum_n|a_n^i|^2=1$ and $\sum_i p_i=1$.
Here, $\rho_I(n)=\sum_{i}p_i|a_n^i|^2$ are the diagonal 
elements of the nuclear--spin density operator.
The Hamiltonian $H_0$ commutes with $\delta h^z$ and thus we find
\begin{equation}\label{Eq:overbar}
\overline{\langle\tau^z(t)\rangle}=\sum_n\rho_I (n)\langle\tau^z(t)\rangle_n,
\end{equation}
which defines the overbar.

We assume that for a large number of nuclear spins $N\gg1$ which are in a superposition of
$\delta h^{z}$-eigenstates $\ket{n}$,  $\rho_{I}(n)$
describes a continuous Gaussian distribution of $\delta h_{n}^{z}$
values, with mean $\overline{\delta h^{z}}$ and variance $\sigma^{2}=
\overline{\left(\delta h^{z}-
\overline{\delta h^{z}}\right)^{2}}$.
In the limit of large $N$ the approach to a Gaussian distribution 
for a sufficiently randomized nuclear system is guaranteed by the
central limit theorem.\cite{Coish:2004a} We perform the continuum limit according to
\begin{eqnarray}
\sum_{n}\rho_{I}(n)f(n) & \to & \int dx
\rho_{I;\overline{x},\sigma}(x)f(x),\label{Eq:continuumlimit}\\ 
\rho_{I;\overline{x},\sigma}(x) & = & \frac{1}{\sqrt{2\pi}\sigma}\label{Eq:gaussian}
\exp\left(-\frac{\left(x-\overline{x}\right)^{2}}
{2\sigma^{2}}\right),\end{eqnarray}
where $x=\delta h_n^z$, $\overline{x}=\overline{\delta h^z}$ and $\sigma^2=\overline{x^2}
-\overline{x}^2$.
The only effect of $\delta b^z$ is to shift the mean value of the Overhauser field 
inhomogeneity to $x_0=\overline{x}+\delta b^z$, whereas the width is left unchanged: 
$\sigma_0=\sigma$. According to this description we obtain 
\begin{eqnarray}
\overline{\langle\tau^z(t)\rangle}&=& \int_{-\infty}^{\infty}dx \rho_{I;x_0,\sigma_0}(x)\left(
f(x)+g(x,t)\right),\\
f(x)&=&\frac{(2x-\omega)^2}{(2x-\omega)^2+(j/2)^2},\\
g(x,t)&=&\frac{(j/2)^2\cos
  \left(2\sqrt{(2x-\omega)^2+(j/2)^2}t\right)}{(2x-\omega)^2+(j/2)^2}. \label{Eq:gterm}
\end{eqnarray}
The second term (Eq.(\ref{Eq:gterm})) vanishes when it is averaged over time and we find
\begin{widetext}
\begin{equation}\label{Eq:convolution}
1-\left\langle\overline{\langle\tau^z\rangle}\right\rangle=\frac{1}{2 \sigma_0\sqrt{2\pi}}
\int^{\infty}_{-\infty}dx\exp\left(-\frac{(x-2 x_0)^2}{8\sigma_0^2}\right)
\frac{(j/2)^2}{(x-\omega)^2+(j/2)^2}.
\end{equation}
\end{widetext}
This integral (a convolution of a Lorentzian and Gaussian) is the well-known Voigt function,
\cite{Armstrong:1967a} and the resulting lineshape is the so-called ``Voigt profile''. The
Voigt function may be expressed as ($\tilde{\omega}=j+4ix_0-2i\omega$)
\begin{eqnarray}
\left\langle\overline{\langle\tau^z\rangle}\right\rangle&=&1-\frac{j}{4 \sigma_0}\sqrt{\frac{\pi}{2}}
\mathrm{Re}\left[\exp\left(\frac{\tilde{\omega}^2}{32 \sigma_0^2}\right)
\mathrm{erfc}\left(\frac{\tilde{\omega}}{4\sqrt{2}\sigma_0}\right)\right],\nonumber \\ 
 & &
\end{eqnarray}
where $\mathrm{erfc}(z)$ is the complementary error function. In the regime where $\sigma_0\ll j$
we may approximate the Lorentzian in the convolution
(Eq.(\ref{Eq:convolution}))by its value at $x=2x_0$ and obtain 
\begin{equation}\label{Eq:voigtlor}
\left\langle\overline{\langle\tau^z\rangle}\right\rangle\approx \frac{(2x_0-\omega)^2}
{(2x_0-\omega)^2+(j/2)^2};\,\,\,\,\,\sigma_0\ll j.
\end{equation}
In this case the resulting resonance has the same FWHM as the Lorentzian, viz. $j$.
On the other hand, if $\sigma_0\gg j$, we may approximate the Gaussian with its value
at $x=\omega$ and thus obtain
\begin{equation}
\left\langle\overline{\langle\tau^z\rangle}\right\rangle \approx
1-\frac{j}{4 \sigma_0}\sqrt{\frac{\pi}{2}} 
\exp\left(-\frac{(2x_0-\omega)^2}{8\sigma_0^2}\right);\,\,\,\,\,\sigma_0\gg j.
\end{equation}
In this regime the width is twice the width $\sigma_0$ of the Gaussian
distribution of the nuclear spin 
states. In order to make a statement about the width of the Voigt profile in general we look
at the peak--to--peak separation $\Delta_V$ of the first derivative of the Voigt profile. 
For a Gaussian 
with a standard deviation of $2 \sigma_0$ we find $\Delta_G=4\sigma_0$ for the peak--to--peak 
separation of the derivative and for a Lorentzian with FWHM of $j$ we have $\Delta_L=j/\sqrt{3}$.
A Pad\'e approximant for $\Delta_V$ in terms of $\Delta_L$ and $\Delta_G$ yields
\cite{Stoneham:1972a}
\begin{equation}\label{Eq:totalwidth}
\Delta_V=\frac{\Delta_G^2+a_1 \Delta_G \Delta_L + a_2 \Delta_L^2}{\Delta_G+a_2 \Delta_L}
\end{equation}
where $a_1=0.9085$, $a_2=0.4621$. This approximation is accurate to better than $0.01 \Delta_V$ 
for all values of $\Delta_L,\Delta_G$. \cite{Stoneham:1972a} A similar formula may also be given 
for the half width at half maximum (HWHM) of the Voigt profile.\cite{Minguzzi:1985a}

\section{State narrowing \label{sub:squeezing}}

The general idea behind state narrowing is that the evolution of the 
two--electron system is dependent on the nuclear spin state and thus knowing 
the evolution of the two--electron system determines the nuclear spin state.
Thus, in this section we describe how the Gaussian superposition
$\rho_{I;\sigma_0,x_0}(x)$ of collective  
nuclear spin eigenstates $\ket{n}$ can be narrowed through a sequence
of measurements performed on a double quantum dot on a time
scale much less than the timescale of 
variation of $\delta h^z$ and for $j \lesssim \sigma_0$. We first
give a general description of how a complete measurement of the lineshape of the Rabi
resonance narrows the Gaussian superposition. Such a complete
measurement of the lineshape consists of many single measurements of
the operator $\tau^z$. In Sec. \ref{sub:singlemeasurement} we
present a detailed analysis of such a complete measurement and in Sec.
\ref{sub:schemes} we discuss different measurement schemes. 

\begin{figure}[h!]
\scalebox{0.8}{\includegraphics{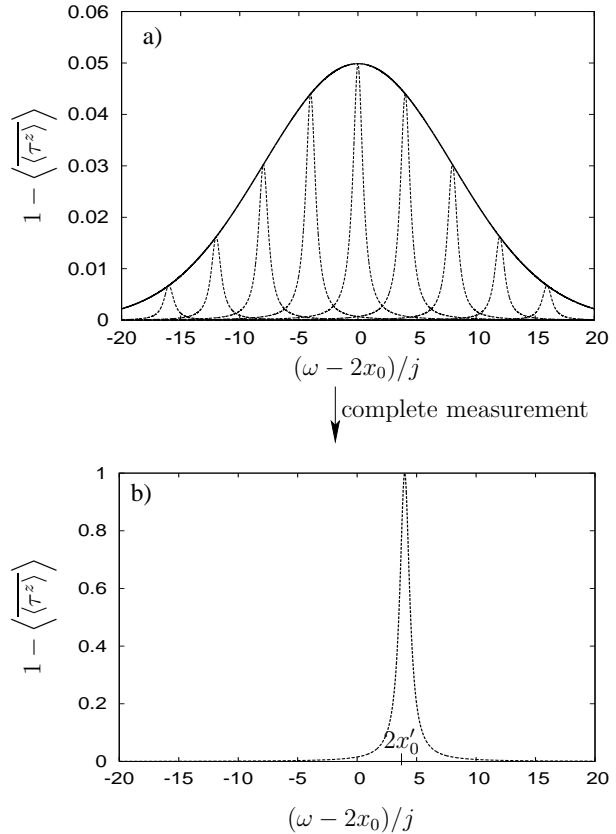}}
\caption{\label{Fig:measurement} a) This figure illustrates the projection obtained through an 
ideal complete measurement of the Rabi--resonance lineshape. All
the different Lorentzian resonances  
corresponding to 
different nuclear spin eigenstates add up to a Gaussian lineshape. 
b) Through a perfect complete
measurement of 
the lineshape of the Rabi resonance, which involves many single measurements of $\tau^z$,
the superposition collapses and we are left with one single Lorentzian 
centered around $2x_0^{\prime}=\Omega_n$, which in general is different from $2x_0$.}
\end{figure}

The operator $\delta h^z$ was defined in Sec. \ref{sec:Rabi} and it
describes the difference in the z-components of total nuclear field in each of
the two dots. The total nuclear field is the result of $N \sim 10^6$
single nuclear spins and thus the eigenvalues of $\delta h^z$
will be highly degenerate. In the limit of large $N$ the spectrum of
$\delta h^z$ is quasi-continuous and the probability density of
eigenvalues of $\delta h^z$ is given by a Gaussian distribution, as
described in Sec. \ref{sub:superposition}.
For such a Gaussian superposition of nuclear spin eigenstates, the lineshape
of the Rabi resonance is given by a Voigt profile, as described in
Sec. \ref{sub:superposition}. This Voigt profile can be seen as a
superposition of Lorentzian lineshapes, where each Lorentzian
results from a nuclear spin eigenvalue $\delta h^z_n$ and is centered
around $\Omega_n=2(\delta h^z_n+\delta b^z)$. In the Voigt profile,
these Lorentzian lineshapes are weighted according to the 
amplitude of the corresponding eigenvalue $\delta h^z_n$ in the
Gaussian-distributed superposition. Through a perfect complete
measurement of the Rabi--resonance lineshape, the superposition
of Lorentzian lineshapes collapses and we are left with one single
Lorentzian (see figure \ref{Fig:measurement}). 
This Lorentzian corresponds to one single eigenvalue of 
$\delta h^z$ and thus the Gaussian distribution has been narrowed to
zero width; the nuclear--spin system is in a state with fixed
eigenvalue $\delta h^z_n$.

In principle, we would need to do infinitely many single measurements in order
to completely measure the lineshape of the Rabi resonance with perfect accuracy,
since each point on this resonance curve is a (time--averaged)
expectation value of the quantum mechanical operator $\tau^z$. Still, we may
perform a finite number $M$ of single measurements (see Sec.
\ref{sub:singlemeasurement}) for each of a set of
driving frequencies $\omega$ and thus obtain the series of expectation
values for different $\omega$ up to some
error. This error depends on $M$. There will then in general be more
than one Lorentzian which can be 
fit (within error) to these
expectation values and thus we would not narrow to zero width. We
would still
have a distribution of nuclear spin eigenstates, but one with smaller
width than before the measurements. 

For such a narrowing through measurement to be successful, the
amplitude $j$ of the oscillating exchange 
$J(t)$ which determines the width of the Lorentzian lineshapes should be
smaller than the width $\sigma_0$ of the Gaussian distribution. Otherwise, the
Rabi resonance would be dominated by the Lorentzian (see 
Eq.(\ref{Eq:voigtlor})) and the method would not result in narrowing
of the nuclear--spin distribution.
The general requirements on the system parameters to narrow the
distribution of nuclear spin eigenvalues are 
\begin{equation}
j,J_0, \sigma_0 \ll x_0;\,\,\,\,\,j \lesssim \sigma_0.
\end{equation}
We note that, unlike in standard ESR, power absorption is not measured
here, but instead the expectation value of the pseudo-spin $\tau^z$, for
instance via a quantum point contact (QPC) nearby one quantum dot (for
a detailed description of the measurement process via such a QPC we
refer the interested reader to Ref. \onlinecite{Engel:2004a}).  To determine 
the expectation value of the pseudo-spin $\tau^z$ many single measurements of 
the pseudo-spin are necessary and we thus proceed to give a detailed description of 
the state narrowing by considering the effect of these single measurements on the
nuclear spin state.

\subsection{Description of state narrowing by consecutive pseudo--spin measurements 
  \label{sub:singlemeasurement}}

In this subsection we describe in detail how a single measurement of
the pseudo-spin $\tau^z$ of the two--electron system affects the
nuclear--spin system. Further, we give a general formula for the
diagonal elements of the nuclear--spin--system density operator in the
continuum limit after $M$ measurements. The sequence of $M$ measurements is referred to as a ``complete measurement''.  

At $t=0$ the two--electron system is initialized to the state
$\ket{+}=\ket{\downarrow\uparrow}$ and we assume that the electron and
the nuclear system are initially factorized. Thus, the total system at $t=0$
is described generally by the following density operator
\begin{equation}
\rho(0)=\rho_e(0)\otimes\rho_I(0)= \ket{+}\bra{+}\otimes
\sum_ip_i\ket{\psi_I^i}\bra{\psi_I^i},
\end{equation}
with nuclear--spin state $\ket{\psi_{I}^i}=\sum_na_n^i\ket{n}$. The
diagonal elements of the nuclear--spin density operator at $t=0$ are
given by $\rho_I(n)=\rho_I(n,0)=\sum_{i}p_i|a_n^i|^2$ and in the
continuum limit we obtain the probability density
$\rho_{I;\overline{x},\sigma}(x)$ for the eigenvalues $\delta h^z_n=x$ as given in
Eq.(\ref{Eq:gaussian}). At time $t_m$ a measurement of the
two--electron system (at driving frequency $\omega$, where $\omega$ is
defined in Eq.(\ref{Eq:joft})) is performed with two
possible outcomes $\ket{+}$ and $\ket{-}$. The diagonal elements of
the nuclear--spin density operator 
after the measurement are given by (see Appendix \ref{app:measurement})
\begin{equation}
\rho_{I}^{(1,\pm)}(n,t_m)=\frac{\rho_I(n,0)}{P^{\pm}(t_m)}
\frac{1}{2} \left(1 \pm \langle\tau^z(t_m)\rangle_n\right),
\end{equation}
where  $\langle\tau^z(t)\rangle_n$ is given by Eq.(\ref{eq:tauz})
and the probabilities $P^{\pm}(t_m)$ to measure $\ket{\pm}$ are
\begin{equation}
P^{\pm}(t_m)=\sum_i\sum_n\frac{1}{2}
\left(1\pm\langle\tau^z(t_m)\rangle_n\right)p_i|a_n^i|^2.  
\end{equation}

In the case where a measurement is performed with a low time
resolution\cite{timeresolution} $\Delta t$, i.e., if $\Delta t \gg 1/j$, the density
operator after the measurement is the time average over the time
interval $\Delta t$ and the cosine term in
$\langle\tau^z(t_m)\rangle_n$ averages out (note that in the case
of a measurement with low time resolution, $t_m$ is arbitrary, as long as $\Delta t$ is
chosen to be large enough). For the rest of this subsection we thus assume\cite{assumption} that
measurements are performed with low time resolution $\Delta t \gg
1/j$. Further, we perform the 
continuum limit and obtain for the probability density of
eigenvalues, i.e., the diagonal part of the density operator in the
continuum limit (with $x=\delta h^z_n+\delta b^z$ and
$\rho_I(x)\equiv\rho_{I;x_0,\sigma_0}(x)$, see Eq.(\ref{Eq:gaussian})): 
\begin{eqnarray}
\rho_{I}^{(1,+,\omega)}(x) &=&
\rho_I(x)(1-L_{\omega}(x))\frac{1}{P^{+}_{\omega}}, \label{Eq:rhoplus}\\
\rho_{I}^{(1,-,\omega)}(x) &=&
\rho_I(x)L_{\omega}(x)\frac{1}{P^{-}_{\omega}}, \label{Eq:rhominus}
\end{eqnarray}
where the probabilities for measuring $\ket{+}$ or $\ket{-}$ are given by
\begin{eqnarray}
P^{+}_{\omega}  &=& \int_{-\infty}^{\infty} dx
\rho_I(x)(1-L_{\omega}(x)), \\
P^{-}_{\omega}  &=& \int_{-\infty}^{\infty} dx
\rho_I(x)L_{\omega}(x),\label{Eq:pminus}
\end{eqnarray}
with
\begin{equation}\label{Eq:lorentzian}
L_{\omega}(x)=\frac{1}{2}\frac{(j/4)^2}{(x-\frac{\omega}{2})^2+(j/4)^2}.
\end{equation}

After the first measurement, the two--electron system is
reinitialized to the state $\ket{+}$ if necessary and a second
measurement is performed. Since the initial density matrix factors
out in the above results, it is clear how to generalize
Eqs.(\ref{Eq:rhoplus}) and (\ref{Eq:rhominus}) 
to the case where $M$ consecutive measurements (without randomization
of the nuclear--spin system in between measurements) are performed: every
time $\ket{+}$ is measured, the diagonal elements $\rho_I(x)$ of the
nuclear density
matrix is  multiplied by $1-L_{\omega}(x)$ and every time
$\ket{-}$ is measured,  $\rho_I(x)$ is multiplied by 
$L_{\omega}(x)$. Thus, we obtain the diagonal elements
$\rho_{I}^{(M,\alpha^{-},\omega)}(x)$ of the nuclear density matrix after $M$ measurements,
of which $\alpha^{-}$ times the measurement outcome was $\ket{-}$ (and
$(M-\alpha^{-})$--times $\ket{+}$):
\begin{equation}\label{Eq:rhofixedfreq}
\rho_{I}^{(M,\alpha^{-},\omega)}(x)=\frac{\rho_I(x)}{Q_{\omega}(M,\alpha^{-})}W_{\omega}(M,\alpha^{-};x).
\end{equation}
Here, $W_{\omega}(M,\alpha^{-};x)$ and the
normalization factor $Q_{\omega}(M,\alpha^{-})$ are given by 
\begin{eqnarray}
W_{\omega}(M,\alpha^{-};x)&=& L_{\omega}(x)^{\alpha^{-}}
(1-L_{\omega}(x))^{M-\alpha^{-}},\\
Q_{\omega}(M,\alpha^{-})&=&\int_{-\infty}^{\infty}dx
\rho_I(x)W_{\omega}(M,\alpha^{-};x).
\end{eqnarray}
The normalization factor $Q_{\omega}(M,\alpha^{-})$ is related to
$P^{\pm}_{\omega}$ through $P^{-}_{\omega}=Q_{\omega}(1,1)$, 
$P^{+}_{\omega}=Q_{\omega}(1,0)$. In the case where
measurements are performed at $m_f$ 
different frequencies, Eq.(\ref{Eq:rhofixedfreq}) generalizes to 
\begin{equation}\label{Eq:rhomultifreq}
\rho_{I}^{(\{M_i\},\{\alpha^{-}_i\},\{\omega_i\})}(x)=\rho_I(x)\prod_{i=1}^{m_f}
\frac{W_{\omega_i}(M_i,\alpha_i^{-};x)}{Q_{\omega_i}(M_i,\alpha^{-}_i)}.
\end{equation}
 The probability density 
 $\rho_{I}^{(\{M_i\},\{\alpha^{-}_i\},\{\omega_i\})}(x)$ after 
$M$ measurements performed at $m_f$ different driving frequencies 
depends on the frequencies $\{\omega_i\} = \{\omega_1, \dots ,\omega_{m_f}\}
$, the number of measurements at each frequency $\{M_i\}=
\{M_1, \dots ,M_{m_f}\}$, and the number of times $\ket{-}$ was measured at
each frequency $\{\alpha^{-}_i\}=\{\alpha^{-}_1, \dots ,
\alpha^{-}_{m_f}\}$. Eq.(\ref{Eq:rhomultifreq}) gives the distribution
of nuclear spin eigenvalues for any sequence of $M$ measurements, i.e.,
without randomization of the nuclear--spin system in between measurements.

\subsection{Measurement schemes\label{sub:schemes}}

In this subsection we describe different
measurement schemes. One main characteristic of the schemes is 
whether we have unconditional evolution of the nuclear--spin density matrix
between measurements (one waits for the nuclear--spin system to rerandomize between
subsequent measurements), or whether we have conditional evolution,
i.e., the nuclear--spin system is assumed to be static between measurements.

\subsubsection{Unconditional scheme}
The simplest scheme is to measure only once at one single driving
frequency $\omega$. If the outcome is $\ket{-}$, the nuclear--spin distribution after the
measurement is given by Eq.(\ref{Eq:rhominus}); the FWHM
($2\sigma_0\sqrt{2\ln 2}\approx 2\sigma_0$) of the
initial distribution will have been narrowed by a factor $\approx
j/4 \sigma_0$ (the nuclear--spin distribution will approximately be a
Lorentzian with FWHM of $j/2$). For $j\ll \sigma_0$ and $\omega=2x_0$, 
the probability $ P^{-}_{\omega}$ to
measure $\ket{-}$ in the first measurement is $P^{-}_{\omega=2x_0} \approx j/6 \sigma_0$
(the exact formula is given in Eq.(\ref{Eq:pminus})). If the 
measurement outcome is $\ket{-}$, we stop measuring. Otherwise, we wait
for the system to rerandomize (in contrast to the conditional schemes) and perform another measurement. This is repeated
until $\ket{-}$ is measured for the first time. On average one
needs to perform $M'\approx 6 \sigma_0/j$ measurements in order to
narrow by a factor of $\approx j/4 \sigma_0$ (we write $M'$ because this number
of measurements should not be confused with the number of measurements
$M$ used above in the case of measurements performed without rerandomization
in between). If the driving frequency
$\omega$ is far from the center $x_0$ of the initial Gaussian distribution,
the number of required measurements increases by a factor of
$\exp((x_0-\omega/2)^2/2\sigma_0^2)$. This
always leads to a narrowed distribution which is
centered around $\omega/2$. Thus, with this
scheme it is possible to choose the center of the nuclear--spin
distribution after the measurement. 
This unconditional measurement scheme is the one which should be
easiest to implement in an experiment since one only needs to measure
once at one single frequency. However, if measurements at several
different frequencies can be performed, a systematic narrowing of the
distribution can be implemented as we show next. 

\subsubsection{Adaptive conditional scheme}\label{subsub:adaptive}

The probability of measuring  $\ket{-}$ in a measurement is
determined by the overlap of the Lorentzian $L_{\omega}(x)$
and the probability density of eigenvalues $\rho_I^{(M,\alpha^{-},\omega)}(x)$
(for the first measurement this probability is $ P^{-}_{\omega}$, which is given in 
Eq.(\ref{Eq:pminus})). Then, if we have the outcome $\ket{-}$ for a
measurement at driving frequency $\omega$, $\rho_I^{(M,\alpha^{-},\omega)}(x)$
as a function of $x$ becomes peaked around $\omega/2$ ( since
$L_{\omega}(x)$ is centered around $x=\omega/2$),
the overlap of the  Lorentzian $L_{\omega}(x)$
and $\rho_I^{(M,\alpha^{-},\omega)}(x)$ increases and therefore the
probability to measure $\ket{-}$ in a 
subsequent measurement also grows. If, on the other hand, we have
outcome $\ket{+}$, the term $1-L_{\omega}(x)$ causes a dip in
$\rho_I^{(M,\alpha^{-},\omega)}(x)$ at $x=\omega/2$, the overlap  of
the  Lorentzian $L_{\omega}(x)$ 
and $\rho_I^{(M,\alpha^{-},\omega)}(x)$ decreases
and thus the probability to measure $\ket{-}$ in a subsequent
measurement with the same driving frequency $\omega$
also decreases. Since it is the measurement outcome $\ket{-}$ that
primarily leads to narrowing, the measurement scheme should maximize the
probability to measure $\ket{-}$. This can be achieved by changing the
driving frequency $\omega$ always in such a way that before each measurement
$L_{\omega}(x)$ and the nuclear--spin 
distribution $\rho_I^{(M,\alpha^{-},\omega)}(x)$ have their maximum at
the same $x$, i.e., set $\omega/2=x_{\mathrm{max}}$, where
$x_{\mathrm{max}}$ is the $x$ for which
$\rho_I^{(M,\alpha^{-},\omega)}(x)$ has a maximum. Thanks to the
adaptive driving frequency $\omega$, the probability
$P^{-}_{\omega}$ to measure $\ket{-}$ is $\approx j/6\sigma_0$ in each measurement until
$\ket{-}$ is measured for the first time. Without adapting, i.e., when
measuring always at the same driving frequency $\omega$, $P^{-}_{\omega}$ decreases, as
explained above (as long as we do not measure $\ket{-}$). After
measuring $\ket{-}$ for the first time,
the probability $P^{-}_\omega$ to measure $\ket{-}$ increases. Every time the measurement
outcome is $\ket{-}$, the distribution
$\rho_I^{(M,\alpha^{-},\omega)}(x)$ is multiplied by $L_{\omega}(x)$
and becomes narrower (since $L_{\omega}(x)^{\alpha^{-}}$ has a FWHM of $(j/2)\sqrt{2^{1/\alpha^{-}}-1}$). However, the measurement outcome $\ket{+}$, for which 
$\rho_I^{(M,\alpha^{-},\omega)}(x)$ is multiplied by
$1-L_{\omega}(x)$, is still more likely and leads to a small widening
of the distribution. Our simulations of this measurement scheme do, however,
show that after $\ket{-}$ has been measured several times, the nuclear spin
distribution is narrowed by more than a factor $j/4\sigma_0$.  

This adaptive scheme
was first proposed in an optical setup by Stepanenko et al. in
Ref. \onlinecite{Stepanenko:2005a}. This scheme requires that
$x_{\mathrm{max}}$ can be calculated (or read from a table) between subsequent measurements
and that the driving frequency $\omega$ can be tuned with a precision that is better than
the width of the nuclear--spin distribution before each measurement. 
For this adaptive scheme (and other conditional schemes)
to work, it is important that the nuclear--spin system does {\it not}
randomize during the course of the 
complete measurement, i.e., the complete measurement must be carried out within a time
that is shorter than the time scale for nuclear spin dynamics.
We thus assume that the nuclear--spin system
(viz. $\delta h^z$) has no internal dynamics 
between the single measurements of $\tau^z(t)$, but only changes due to the
measurements performed on the two--electron system, i.e., due to single
measurements of $\tau^z(t)$.
We expect $\delta h^z$ to vary on the time scale of nuclear spin
diffusion out of the dot, which is on the order of seconds for nuclear
spins surrounding donor impurities in GaAs.\cite{Paget:1982a} 
However, there may be other sources of nuclear spin dynamics (see also Appendix 
\ref{App:chemicalshift}).

\begin{figure}[h!]
\scalebox{0.42}{\includegraphics{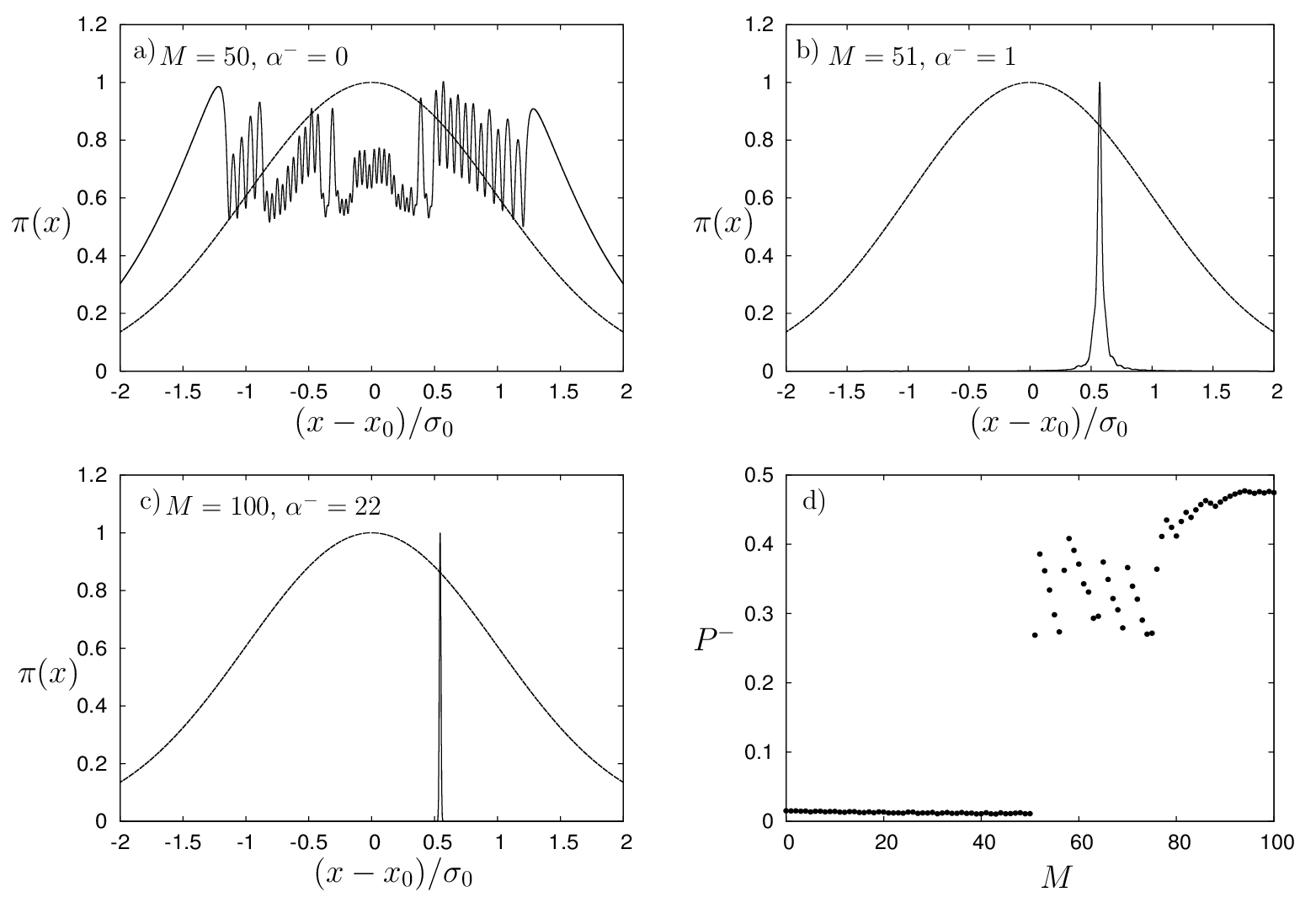}}
\caption{\label{fig:adaptive}In this figure we show a typical\cite{typical} sequence of
  the rescaled probability density of eigenvalues
  $\pi(x)=\rho_I^{(\{M_i\},\{\alpha^-_i\},\{\omega_i\})}(x)/
  \mathrm{max}\left(\rho_I^{(\{M_i\},\{\alpha^-_i\},
      \{\omega_i\})}(x)\right)$ for the adaptive conditional scheme.       
  Here, $\rho_I^{(\{M_i\},\{\alpha^-_i\},\{\omega_i\})}(x)$ is given
  in Eq.(\ref{Eq:rhomultifreq}). We have 
  $x=\delta h^z_n+\delta b^z$, $j/\sigma_0=1/10$ and in a)--c) the
  initial Gaussian distribution (with FWHM $2\sigma_0\sqrt{2\ln
    2}\approx 2\sigma_0$) is plotted for reference.
 a) Up to $M=50$ measurements
the outcome is never $\ket{-}$ and thus each measurement ``burns a
hole'' into the distribution where it previously had its maximum. b)
In the $51^{\mathrm{st}}$ measurement the outcome is $\ket{-}$
which leads to a narrowed distribution of nuclear spin eigenvalues
(peak centered at $\approx 0.5$) with a FWHM that is reduced by a factor
$\approx j/4 \sigma_0$. c) Adapting the 
driving frequency $\omega$ to this peak, i.e., setting $\omega/2=x_{\mathrm{max}}$ in subsequent
measurements, leads to further narrowing
every time $\ket{-}$ is measured. In
this example the final FWHM is $\approx\sigma_0/100$, i.e., the
distribution has been narrowed by a 
factor $\approx j/10\sigma_0$. d) The probability $P^-$ to 
measure $\ket{-}$ jumps up after the $51^{\mathrm{st}}$ measurement and after
$\ket{-}$ is measured several more times, this probability saturates close to $1/2$.} 
\end{figure}

In figure \ref{fig:adaptive} we show a typical\cite{typical} sequence of nuclear
spin distributions for the adaptive scheme with total number of
measurements $M=100$ and $j/\sigma_0=1/10$. We see ( figure
\ref{fig:adaptive} (a)) that up to $M=50$
the measurement outcome is never $\ket{-}$ and thus each measurement
``burns a hole'' into the distribution where it previously had its
maximum. In the $51^{\mathrm{st}}$ measurement (figure
\ref{fig:adaptive}(b)) the outcome is $\ket{-}$, which
narrows the distribution by a factor of $\approx j/4\sigma_0$. Adapting the driving frequency
$\omega$ to
this peak, i.e., setting $\omega/2=x_{\mathrm{max}}$ in subsequent
measurements, leads to further narrowing, i.e., to a total narrowing
by more than a factor $j/4\sigma_0$ 
(figure \ref{fig:adaptive}(c)). In this example we have 
$\alpha^-=22$ after $M=100$ measurements and the final FWHM is
$\approx \sigma_0/100$, i.e., the distribution has been narrowed by a
factor $\approx j/10\sigma_0$.
In figure \ref{fig:adaptive}(d) the probability $P^-$ to
measure $\ket{-}$ before each measurement is shown. After the first
time $\ket{-}$ is measured, $P^-$ jumps up and after several more times
$\ket{-}$ was measured, it saturates close to $1/2$. $P^-$ is
a good signature of the distribution's width. As the width of
the distribution goes to zero, $P^-$ approaches $1/2$. 
This adaptive conditional scheme is more intricate than the
unconditional scheme, but allows one to narrow by more than a factor $j/4\sigma_0$.

\subsubsection{Other conditional schemes}
Other possible measurement schemes involve measurements at several
frequencies, as in the adaptive scheme. One may either choose a fixed
number of frequencies within one or two $\sigma_0$ and measure several
times at each frequency (without randomization between the
measurements) or sweep the frequency, i.e., measure only once at each
frequency but vary the frequency only in small steps. Based on
numerical simulations of these schemes, we find that the typical
number of measurements to narrow by a factor of $j/\sigma_0$ is
greater than in the adaptive or the unconditional (single--frequency)
schemes. 

\subsubsection{Time-domain measurement scheme}

We note that when a complete measurement of one of the correlators
discussed in Sec.
\ref{sec:SzZero} is performed with perfect resolution in time and
perfect accuracy, this would also determine the state of the nuclear
spin system and thus narrow the distribution of nuclear spin
states. This is because the frequency of the oscillating correlators is given by
$\sqrt{J^2+4(\delta h^z_n)^2}$ and thus measuring the frequency of the correlator determines
the eigenvalue $\delta h^z_n$ of the nuclear--spin system. However, it
may be possible to perform a weak measurement of the decay of the
correlators and thus also to see the prolongation of the decay after
applying a narrowing scheme. To understand in detail the effect of
measurements in the time domain, further study is required. 
Narrowing through measurement of the correlators is a time--domain
measurement. In contrast, the narrowing schemes we have proposed
above are frequency--domain measurements. If the frequency resolution
is better than the time resolution, our method would most likely be
more suitable.

\section{\label{sec:SzZero} Correlation Functions in the $S^{z}=0$ Subspace}

In this section we investigate the Hamiltonian $H_0$ of Eq. (\ref{Hamiltonian}) with static
exchange coupling $J$. Using this Hamiltonian we wish to calculate correlation functions 
for several observables in the subspace of zero total spin in the $z$--direction. In our
previous  work \cite{Coish:2005a} we calculated the time evolution of a particular
correlator involving the states $\ket{S}$ and $\ket{T_0}$. However, there are four additional 
independent correlators
involving the $x$ and $y$ components of pseudo--spin which require a separate calculation. 
Quite surprisingly, it will turn out that these correlators have different decay behavior 
in time.
The correlators we calculate here show the 
decoherence properties of the pseudo--spin states under the influence of the hyperfine 
interaction. There 
may be additional sources of decoherence which we do not consider here,
such as orbital dephasing, corrections to the effective Hamiltonian, \cite{Coish:2005a}
the coupling of the QPC to the dot spins, \cite{Borhani:2005a} etc.
The results of this section will help to give requirements on the
parameters of the system in order to  
initialize in the state $\ket{\uparrow\downarrow}$ and to assess the
fidelity of a $\sqrt{\mathrm{SWAP}}$ 
operation with static $J$ (see Sec. \ref{sec:SWAP}). 

Diagonalizing $H_{0}$ gives the following eigenvalues and eigenvectors
\begin{eqnarray}
E_{n}^{\pm} & = & \frac{J}{2}\pm\frac{1}{2}\sqrt{J^{2}
+\Omega_n^{2}},
\label{eq:LargeBEigenvals}\\
\ket{E_{n}^{\pm}} & = & \frac{(\Omega_n/2)\ket{S}+E_{n}^{\pm}\ket{T_{0}}}
{\sqrt{\left(E_{n}^{\pm}\right)^{2}
+\left(\Omega_n/2\right)^{2}}}\otimes\ket{n},
\end{eqnarray}
where again $\ket{n}$ is an eigenstate of the operator $\delta h^{z}$
with $\delta h^z \ket{n}=\delta h_{n}^{z}\ket{n}$.
At $t=0$ we fix the electron system in an arbitrary superposition of $\ket{T_0}$ and $\ket{S}$
\begin{equation}
\ket{\psi_e(t=0)}= \ket{A};\,\,\, \ket{A}=\cos \frac{\theta_A}{2} \ket{S}
+e^{i \varphi_A}\sin \frac{\theta_A}{2} \ket{T_0}.
\end{equation}
The nuclear--spin system is again in a general state
 (see Sec. \ref{sub:superposition}).
As will be shown in Sec. \ref{sec:SWAP}, it is possible, in principle, to initialize to an
arbitrary state in the subspace spanned by $\ket{T_0}$ and $\ket{S}$.
The probability to find the electron spins in a state $\ket{B}$ at $t>0$ is given by the 
correlation function:
\begin{equation}
C_{BA}(t)=\sum_n\rho_I(n)\left| \bra{n}\otimes \bra{B}e^{-iH_0t}\ket{A} \otimes \ket{n}\right|^2,
\end{equation}
where $\rho_I(n)=\sum_i p_i |a^i_n|^2$. The correlation function has the following symmetry:
$C_{BA}(t)=C_{AB}(-t)$, and if $\ket{B}$ and $\ket{D}$ are orthogonal states we have
$C_{BA}(t)=1-C_{DA}(t)$. Further, we may decompose $C_{BA}(t)$ into the sum of a time-independent 
term $\overline{C^n_{BA}}$ and an interference term $C^{\mathrm{int}}_{BA}(t)$:
\begin{equation}
C_{BA}(t)=\overline{C^n_{BA}}+C^{\mathrm{int}}_{BA}(t),
\end{equation}
where the overbar is defined in Eq. (\ref{Eq:overbar}). 

We have further $C^n_{BA}=C_{BA}(\delta h_n^z)=C_{BA}(x)$. Performing the continuum limit as
described in Eq. (\ref{Eq:continuumlimit}) we obtain for the correlation function
\begin{eqnarray}
C_{BA}(t)&=&\int_{-\infty}^{\infty}dx \rho_{I;\sigma_0,x_0}(x) \left(C_{BA}(x)
+C^{\mathrm{int}}_{BA}(x,t)\right)
\label{correlatorintegral}\nonumber\\
& &\\
&=& C_{BA}^{\infty}+C_{BA}^{\mathrm{int}}(t).
\end{eqnarray}
Here, $C_{BA}^{\infty}$ is the assymptotic value of the correlator
$C_{BA}(t)$ for $t\rightarrow \infty$. 

We have calculated correlation functions for the following states: $
\ket{S}\to\ket{\tau^{z}=-1},\, 
\ket{T_{0}}\to\ket{\tau^{z}=+1},$ $ \ket{X}\to\ket{\tau^{x}=+1}=\frac{1}{\sqrt{2}}
\left(\ket{T_0}+\ket{S}\right),\,\ket{Y}\to\ket{\tau^{y}=+1}=\frac{1}{\sqrt{2}}
\left(\ket{T_{0}}+i\ket{S}\right)$. The frequency in the interference term is
always given by $s(x)=\sqrt{J^2+4x^2}$. In Table
\ref{CorrelatorIntegrands} we list the  integrands
according to the  notation in Eq. (\ref{correlatorintegral}). 
From the Heisenberg equation of motion we find $\frac{d \tau^x}{dt}=-J
\tau^y$, which leads to relations for the correlators. In the notation used in Table 
\ref{CorrelatorIntegrands} we obtain $\frac{d
  C_{XX}}{dt}=-J\left(C_{YX}-\frac{1}{2}\right)$, 
which is satisfied by the results shown in Table
\ref{CorrelatorIntegrands}. 
Similar relations can be derived for the other correlators and used to
check the results in Table \ref{CorrelatorIntegrands}. 
\begin{table}[!h]
\begin{tabular}{|r||c|c|} \hline 
$C_{BA}(t)$ & $C_{BA}(x)$& $C_{BA}^{\mathrm{int}}(x,t)$\\ \hline \hline
$C_{T_{0}S}(t)$ & $\left. \frac{2 x^2}{s(x)^2}\right.$ &
$\left.-\frac{2 x^2}{s(x)^2}\cos(s(x)t)\right.$ \\ \hline 
$C_{T_{0}X}(t)$ & $ \frac{1}{2}+\frac{J x}{s(x)^2}$ & $-\frac{J x}{s(x)^2}\cos(s(x)t)$ \\ \hline
$C_{T_{0}Y}(t)$ & $ \frac{1}{2}$ & $ \frac{x}{s(x)}\sin(s(x)t)$ \\ \hline
$C_{YX}(t)$ & $\frac{1}{2}$ & $ \frac{J}{2 s(x)}\sin(s(x)t)$ \\ \hline
$C_{YY}(t)$ & $\frac{1}{2}$ & $\frac{1}{2}\cos(s(x)t)$ \\ \hline
$C_{XX}(t)$ & $\frac{1}{2}+\frac{2 x^2}{s(x)^2}$ & $\frac{J^2}{2 s(x)^2}\cos(s(x)t)$ \\ \hline
\end{tabular}
\caption{\label{CorrelatorIntegrands} Functions $C_{BA}(x)$ and $C_{BA}^{\mathrm{int}}(x,t)$
according to the notation of Eq. (\ref{correlatorintegral}) for different correlators (with
$s(x)=\sqrt{J^2+4x^2}$). $C_{XX}(t)$ is a linear combination of other correlators.}
\end{table}
We see that $C_{XX}(t)$ is a linear combination of other correlators: $C_{XX}(t)=C_{YY}(t)
+C_{T_0 S}(t)$. For $C_{T_0X}$ and $C_{T_0Y}$ the interference term is an odd function in $x$.
Thus, the time dependence vanishes for $x_0=0$ and we have $C_{T_0X}=C_{T_0Y}=1/2$ for 
all $t$. In general, the integral in Eq. (\ref{correlatorintegral}) is difficult to solve 
exactly. Thus, we concentrate on several interesting limits. We illustrate this for the 
case of $C_{YX}(t)$ and give results for the other correlators. We have
\begin{eqnarray}
C_{YX}(t)& = &\frac{1}{2}+\mathrm{Im}\left[\tilde{C}_{YX}^{\mathrm{int}}\right],\\
\tilde{C}_{YX}^{\mathrm{int}}&=&\int_{-\infty}^{\infty}\rho_{I;\sigma_0,x_0}(x)
\frac{J}{2 s(x)}e^{is(x)t}.
\end{eqnarray}
In the regime of $|x_0| \gg \sigma_0$ the main contribution to the integral comes from a 
narrow region around $x_0$ and we may approximate $\frac{J}{2s(x)}
\approx \frac{J}{2 \omega_0}$ where $\omega_0=s(x_0)$ and in the frequency term 
$s(x)\approx \omega_0 + \frac{4x_0}{\omega_0}(x-x_0)+\dots$ . 
For this to be a good approximation, we require $\frac{2J^2}{w_0^3}(x-x_0)^2 t\ll1$. 
We use $(x-x_0)^2\approx \sigma_0^2$ and thus obtain for the correlator and the range 
of validity in this limit
\begin{eqnarray}\label{YXx0>>sigma}
C_{YX}^{\mathrm{int}}(t)& = &\frac{J}{2\omega_0}
e^{-\frac{1}{2}\left(\frac{t}{t_0^{\prime\prime}}\right)^2}\sin(\omega_0 t), \\
& &t_0^{\prime\prime}=\frac{\omega_0}{4 |x_0|\sigma_0},\,\,\, \omega_0=\sqrt{J^2+4x_0^2}, \\
& &|x_0|\gg \sigma_0,\,\,\,t \ll \frac{(J^2+4 x_0^2)^{3/2}}{2J^2\sigma_0^2}.
\end{eqnarray}
The results for the other correlators are (with the same range of validity)
\begin{eqnarray}
C_{T_0 S}^{\mathrm{int}}(t)&=&-\frac{2 x_0^2}{\omega_0^2}
e^{-\frac{1}{2}\left(\frac{t}{t_0^{\prime\prime}}\right)^2}\cos(\omega_0 t),\\
C_{T_0 X}^{\mathrm{int}}(t)&=&-\frac{J x_0}{\omega_0^2}
e^{-\frac{1}{2}\left(\frac{t}{t_0^{\prime\prime}}\right)^2}\cos(\omega_0 t)\label{T0Xx_0>>sigma},\\
C_{T_0 Y}^{\mathrm{int}}(t)&=&\frac{x_0}{\omega_0}
e^{-\frac{1}{2}\left(\frac{t}{t_0^{\prime\prime}}\right)^2}\sin(\omega_0 t),\\
C_{YY}^{\mathrm{int}}(t)&=&\frac{1}{2}
e^{-\frac{1}{2}\left(\frac{t}{t_0^{\prime\prime}}\right)^2}\cos(\omega_0 t).
\end{eqnarray}
In this limit we obtain a Gaussian decay for all correlators on a time scale 
$t_0^{\prime\prime}=\frac{\omega_0}{4 |x_0|\sigma_0}$ which grows with the absolute 
value of the exchange coupling $|J|$ and with $1/\sigma_0$. The long--time saturation 
value is $1/2$ for $C_{YX}$.
For some of the other correlators we find non--trivial parameter--dependent saturation values. 
In the limit of $|x_0|\gg \sigma_0$ we obtain these correlators by the same approximation as 
for the interference term, i.e. we set $C_{BA}(x)=C_{BA}(x_0)$ and obtain
\begin{eqnarray}
C_{T_0 S}^{\infty}&=& \frac{2 x_0^2}{J^2 +4x_0^2};\,\,\,\,\,|x_0| \gg\sigma_0,\\
C_{T_{0}X}^{\infty}&=&\frac{1}{2}+\frac{J x_0}{J^2+4x_0^2};\,\,\,\,\,|x_0| \gg\sigma_0,\label{T0Xinfty}\\
C_{T_0Y}^{\infty}&=&C_{YX}^{\infty}=C_{YY}^{\infty}=\frac{1}{2}.
\end{eqnarray}
For large $J$ the saturation value is quadratic in $x_0/J$ for $C_{T_0S}$ and linear for 
$C_{T_0X}$. The saturation value for $C_{T_0S}$ goes to zero for $|J|\gg|x_0|$ and for $C_{T_0X}$
approaches $1/2$. $C_{T_0X}^{\infty}$ reaches extrema equal to $\frac{1}{2}
+\frac{1}{4}\mathrm{sign}(J x_0)$ for $|J|=2|x_0|$. \\

Next we consider Eq. (\ref{correlatorintegral}) for $|J|\gg \mathrm{max}(|x_0|,\sigma_0)$ and find
\begin{eqnarray}
s(x)=\sqrt{J^2+4x^2}&\approx & |J|+\frac{2x^2}{|J|},\label{frequencyapprox}\\
\frac{J}{2 s(x)}=\frac{J}{2 \sqrt{J^2+4x^2}}& \approx & 
\mathrm{sign}(J)\left(\frac{1}{2}-\frac{x^2}{J^2}\right).\label{secondapprox} 
\end{eqnarray}
For Eq. (\ref{frequencyapprox}) we have the additional requirement that
$t\ll \frac{|J|^3}{2 \mathrm{max}(x_0^4,\sigma_0^4)}$. Under these approximations we find the
following result:
\begin{widetext}
\begin{eqnarray}\label{correlatorlargeJ}
\tilde{C}_{YX}^{\mathrm{int}}(t)&=&\mathrm{sign}(J)\left(\frac{1}{2}\xi(t)-
\frac{\sigma_0^2}{J^2}\xi^3(t)-\frac{x_0^2}{J^2}\xi^5(t)\right)
\exp\left(i|J|t-\frac{x_0^2}{2 \sigma_0^2}\left(1-\xi^2(t)\right)\right),\\
& &\xi(t)=\left(1-i\frac{t}{t_0^{\prime}}\right)^{-1/2},\,\,\,
t_0^{\prime}=\frac{|J|}{4 \sigma_0^2},\,\,\,  
|J|\gg \mathrm{max}(|x_0|,\sigma_0),\,\,\, t\ll\frac{|J|^3}{2 \mathrm{max}(x_0^4,\sigma_0^4)}.
\end{eqnarray}
\end{widetext}
At short times we expand $\xi^2(t)\sim 1+i\frac{t}{t_0^{\prime}}
-\left(\frac{t}{t_0^{\prime}}\right)^2$. Keeping only lowest order in $t/t_0^{\prime}$
 in the prefactor and second order in the frequency term we obtain
\begin{eqnarray}
C_{YX}^{\mathrm{int}}(t)&=&\mathrm{sign}(J)\frac{1}{2}
e^{-\frac{1}{2}\left(\frac{t}{t_0^{\prime\prime}}\right)^2}
\sin\left(\omega_0^{\prime}t\right)\label{YXlargeJsmallt},\\
& & t_0^{\prime\prime}\approx \frac{|J|}{4|x_0|\sigma_0},\,\,\, 
\omega_0^{\prime}=|J|+\frac{2(x_0^2+\sigma_0^2)}{|J|},\\
& & t\ll t_0^{\prime}=\frac{|J|}{4\sigma_0^2},\,\,\, |J|\gg \mathrm{max}(|x_0|,\sigma_0).
\end{eqnarray}
The $|x_0|\gg \sigma_0$ limit of this result agrees with the $|J|\gg |x_0|$ limit of 
Eq. (\ref{YXx0>>sigma}). Again, we have a Gaussian decay on the same time scale 
$t_0^{\prime\prime}$ as in 
Eq. (\ref{YXx0>>sigma}) ($\omega_0=\sqrt{J^2+4x_0^2}\sim |J|$ for $|J|\gg |x_0|$). One
interesting feature of this correlator is the fact that there is a change of phase by $\pi$ when 
the sign of the exchange coupling $J$ changes. This feature offers the possibility of measuring 
$J$ even for small values of $J$ through a measurement of this correlator. 
We also list the other correlators in this regime:
\begin{eqnarray}
C_{T_0S}^{\mathrm{int}}(t)&=&-\frac{2(x_0^2+\sigma_0^2)}{J^2}
e^{-\frac{1}{2}\left(\frac{t}{t_0^{\prime\prime}}\right)^2}\cos(\omega_0^{\prime}t),
\label{T0SlargeJsmallt}\\
C_{T_0X}^{\mathrm{int}}(t)&=&-\frac{x_0}{J}e^{-\frac{1}{2}\left(\frac{t}{t_0^{\prime\prime}}\right)^2}
\cos(\omega_0^{\prime}t)\label{T0XlargeJsmallt},\\
C_{T_0Y}^{\mathrm{int}}(t)&=&\frac{x_0}{|J|}e^{-\frac{1}{2}\left(\frac{t}{t_0^{\prime\prime}}\right)^2}
\sin(\omega_0^{\prime}t)\label{T0YlargeJsmallt},\\
C_{YY}^{\mathrm{int}}(t)&=&\frac{1}{2}e^{-\frac{1}{2}\left(\frac{t}{t_0^{\prime\prime}}\right)^2}
\cos(\omega_0^{\prime}t)\label{YYlargeJsmallt}.
\end{eqnarray}
Finally, we are also interested in the behavior for large t. Thus, we
expand Eq. (\ref{correlatorlargeJ}) for large times $\xi(t\gg t_0^{\prime})
\sim e^{i\pi/4}\sqrt{t_0^{\prime}/t}$ and obtain
\begin{eqnarray}
C_{YX}^{\mathrm{int}}(t)&\sim&\mathrm{sign}(J)e^{-\frac{x_0^2}{2\sigma_0^2}}\frac{\sqrt{|J|}
\sin(|J|t+\frac{\pi}{4})}{4\sigma_0 t^{\frac{1}{2}}}\label{YXlargeJlarget},\\
& &t\gg t_0^{\prime}=\frac{|J|}{4\sigma_0^2}, |J|\gg \mathrm{max}(|x_0|,\sigma_0).
\end{eqnarray}
For the other correlators we find
\begin{eqnarray}
C_{T_0S}^{\mathrm{int}}(t)&\sim& -e^{-\frac{x_0^2}{2\sigma_0^2}}\frac{\cos(|J|t+\frac{3\pi}{4})}
{4 \sigma_0 \sqrt{|J|}\,t^{\frac{3}{2}}},\\
C_{T_0X}^{\mathrm{int}}(t)&\sim& -\mathrm{sign}(J)e^{-\frac{x_0^2}{2\sigma_0^2}}
\frac{x_0\sqrt{|J|}\cos(|J|t+\frac{3\pi}{4})}{8 \sigma_0^3 t^{\frac{3}{2}}}
\label{T0XlargeJlarget},\\
C_{T_0Y}^{\mathrm{int}}(t)&\sim& e^{-\frac{x_0^2}{2\sigma_0^2}} 
\frac{x_0\sqrt{|J|}\sin(|J|t+\frac{3\pi}{4})}{8 \sigma_0^3 t^{\frac{3}{2}}},\\
C_{YY}^{\mathrm{int}}(t)&\sim&e^{-\frac{x_0^2}{2\sigma_0^2}}\frac{\sqrt{|J|}\cos(|J|t+\frac{\pi}{4})}
{4\sigma_0 t^{\frac{1}{2}}}\label{YYlargeJlarget}.\\
\end{eqnarray}
Thus, the transverse components of the pseudo-spin have a slower decay ($\sim t^{-1/2}$)
than the longitudinal component ($\sim t^{-3/2}$). This results from the fact that the 
Hamiltonian only has fluctuations along only one direction.

\section{\label{sec:SWAP} Analysis of $\sqrt{\mathrm{SWAP}}$}

In this section we analyze the $\sqrt{\mathrm{SWAP}}$ gate using the
correlation functions derived in the previous section, i.e., we analyze
the $\sqrt{\mathrm{SWAP}}$ gate taking into account the
hyperfine--induced decoherence. 
The $\sqrt{\mathrm{SWAP}}$ gate and single--qubit operations can be
used to perform the quantum XOR gate (CNOT) which, in combination with
single--qubit operations, is sufficient for  universal quantum
computation. \cite{Loss:1998a,Barenco:1995a} In
Ref. \onlinecite{Petta:2005b} implementation of $\sqrt{\mathrm{SWAP}}$
has been demonstrated. However, in these experiments there was a
contrast reduction of $\sim 40\%$. Here we show that taking into
account hyperfine induced decoherence, still near--unit fidelity can be
obtained for this operation.

The Hamiltonian of Eq. (\ref{Hamiltonian}) induces unitary time
evolution on the states of the system:
$\ket{\psi(t)}=U(t)\ket{\psi(0)}$ with $U(t)=T \exp(-i\int_0^t
H(t^{\prime})dt^{\prime})$. We assume that $J$ and $x_0$ can be
switched adiabatically \cite{Requist:2005a} on a time scale that is
much  shorter than the time required 
for the gate operation and thus the time evolution operator at time
$\tau_s$ has the form
\begin{equation}
U_s=\exp\left(-i \tau_s H\right).
\end{equation}
In a Bloch--sphere picture this operator induces a
rotation about an axis in the plane spanned by eigenstates of $\tau^x$
and $\tau^z$, $\ket{X}=\ket{\uparrow\downarrow}$ and
$\ket{S}=(\ket{\uparrow\downarrow}-\ket{\downarrow\uparrow})/\sqrt{2}$.\cite{Levy:2002a}    
The axis of rotation is determined by the parameters $J$ and $x_0$.
Through such an operation any state may be rotated into any other
state on the Bloch sphere. 
Thus, it is possible to rotate from $\ket{S}$ to any initial state in 
the subspace of $S^z=0$ by a single operation. This is important since
initialization to the singlet is feasible by preparing a ground--state
singlet with both electrons on the same dot and then changing the
bias. \cite{Petta:2005b} We now investigate initialization to the
state $\ket{X}$ taking into account hyperfine--induced decoherence. 
The scheme we propose here is
different from the one used in Ref. \onlinecite{Petta:2005b}, where
adiabatic passage from the singlet to the
$\ket{\uparrow\downarrow}$--state is used. Our scheme requires control
of $x_0$. We assume the system to be in the singlet state $\ket{S}$ at
$t=0$ and then switch $J$ and $x_0$ such that $J=-2x_0$ and $|x_0|\gg
\sigma_0$. In a Bloch--sphere picture, this corresponds to a
rotation about an axis that halves the angle between $\ket{S}$ and
$\ket{X}$. Since $C_{XS}(t)=C_{SX}(-t)=1-C_{T_0X}(-t)$ we 
have, for the above choice of parameters, according to
Eqs. (\ref{T0Xx_0>>sigma}) and (\ref{T0Xinfty}): 
\begin{eqnarray}
C_{XS}(t)&=&\frac{1}{2}+\frac{1}{4}\left(1-\cos(\sqrt{2}|J|t)
e^{-\frac{1}{2}\left(\frac{t}{t_0^{\prime\prime}}\right)^2}\right),\\
& &J=-2x_0,\,\,\,|x_0|\gg \sigma_0,\\
& &t_0^{\prime\prime}=\frac{1}{\sqrt{2}\sigma_0},\,\,\, t\ll
\frac{(J^2+4x_0^2)^{3/2}}{2J^2x_0^2}.
\end{eqnarray}
This correlator reaches its maximum for $\sqrt{2}|J|t=\pi$, i.e., at  
$\tau_s=\frac{\pi}{\sqrt{2}|J|}$. The time scale for the Gaussian
decay is $t^{\prime\prime}=\frac{1}{\sqrt{2}\sigma_0}$. 
To approach unit fidelity we therefore require $|J|\gg \sigma_0$,
which is the case in the range of validity of the above correlator
since $|x_0|\gg \sigma_0$ and $J$ and $x_0$ are of the same order. At
$t=\tau_s$ we switch $J$ to zero and since $\ket{X}\otimes \ket{n}$ 
is an eigenstate of the remaining Hamiltonian, the system remains in
this product state, untouched by decoherence induced via the nuclear
spins. This scheme thus provides a way to initialize the double
quantum dot system to the state
$\ket{X}=\frac{1}{\sqrt{2}}(\ket{T_0}+\ket{S})=\ket{\uparrow\downarrow}$,
where arrows denote the z-component of the electron spin in each dot. 
In the same way, it is also possible to initialize in the state
$\ket{-X}= \ket{\tau^x=-1} =\frac{1}{\sqrt{2}}(\ket{T_0}-\ket{S})
=\ket{\downarrow\uparrow}$ by switching to $J=2x_0$. 

It was already proposed in Ref. \onlinecite{Loss:1998a} to
implement the $\sqrt{\mathrm{SWAP}}$ gate by pulsing the exchange
interaction $J$ between the two dots. Here we give a detailed
analysis of the $\sqrt{\mathrm{SWAP}}$ gate taking into account
hyperfine--induced decoherence.  
 
The SWAP operation acts on the basis of the two--electron system as:
$\ket{\downarrow\downarrow}\to\ket{\downarrow\downarrow}, 
\ket{\downarrow\uparrow}\to\ket{\uparrow\downarrow}, 
\ket{\uparrow\downarrow}\to\ket{\downarrow\uparrow},
\ket{\uparrow\uparrow}\to\ket{\uparrow\uparrow}$. The SWAP is an
operation that acts only on the subspace of $S^z=0$  and leaves the
states $\ket{\uparrow\uparrow}$ and $\ket{\downarrow\downarrow}$
unchanged. In the system we consider this is naturally implemented
through the large Zeeman splitting that separates
$\ket{\uparrow\uparrow}$ and $\ket{\downarrow\downarrow}$ from the
singlet and the $S^z=0$ triplet. In order to analyze the SWAP in
the $S^z=0$ subspace we consider the regime of $|J|\gg
\mathrm{max}(x_0,\sigma_0)$. The correlator $C_{-X,X}(t)$ gives the
probability of being in the state $\ket{-X}=\ket{\downarrow\uparrow}$
for a system initialized in $\ket{X}=\ket{\uparrow\downarrow}$. Due to
the symmetry relations for the correlation functions we have
$C_{-X,X}(t)=1-C_{XX}(t)=1-C_{YY}(t)-C_{T_0S}(t)$ and thus find (using
Eqs. (\ref{T0SlargeJsmallt}) and  (\ref{YYlargeJsmallt}) and neglecting
terms of order $(\sigma_0^2+x_0^2)/J^2$),
\begin{eqnarray}
& &C_{-X,X}(t)=1-C_{XX}(t)\approx \frac{1}{2}
-\frac{1}{2}e^{-\frac{1}{2}\left(\frac{t}{t_0^{\prime\prime}}\right)^2}
\cos(|J|t)\nonumber,\\
& & \\
& &t_0^{\prime\prime}=\frac{|J|}{4\sigma_0|x_0|},\,\,\,|J|\gg
\mathrm{max}(|x_0|,\sigma_0), \,\,\,
t \ll t_0^{\prime}=\frac{|J|}{4 \sigma_0^2}.\nonumber \\
& & 
\end{eqnarray}
We obtain the maximum value for this correlator when
$\tau_s=\frac{\pi}{|J|}$. The Gaussian has a decay time of
$t_0^{\prime\prime}=\frac{|J|}{4\sigma_0|x_0|}$, so for $x_0 \to 0$
the Gaussian decay is negligible and we obtain unit fidelity for this
SWAP operation $\ket{\uparrow\downarrow}\to\ket{\downarrow\uparrow}$
up to a global phase factor (which is not visible in the correlator).

From the SWAP operation it is only a small step towards the
$\sqrt{\mathrm{SWAP}}$ which we obtain when we let the system evolve
with the same parameter values but for only half the time. 
Starting in the state $\ket{X}$ we obtain $\ket{Y}$ after applying a
$\sqrt{\mathrm{SWAP}}$. For large $|J|$ we find for the correlator
$C_{YX}$ in the limit $x_0 \to 0$ 
\begin{eqnarray}
& &C_{YX}(t)=\frac{1}{2}+\mathrm{sign}(J)\frac{1}{2}
e^{-\frac{1}{2}\left(\frac{t}{t_0^{\prime\prime}} 
\right)^2}\sin(|J|t),\\
& &t_0^{\prime\prime}=\frac{|J|}{4\sigma_0|x_0|}, \,|J|\gg
\mathrm{max}(|x_0|,\sigma_0),\, 
t \ll t_0^{\prime}=\frac{|J|}{4 \sigma_0^2}.\nonumber \\
& &
\end{eqnarray}
Here again the time scale of the Gaussian decay is
$\frac{|J|}{4\sigma_0|x_0|}$ and approaches infinity for $x_0 \to
0$. The time during which we have to operate with these values of 
the parameters $J$ and $x_0$ is now $\tau_s=\frac{\pi}{2|J|}$. 
Our calculations show that for the time during which $J$ is pulsed
high there is a regime in which unit fidelity may be approached. The
reduced visibility in the experiment \cite{Petta:2005b} may be due 
to several reasons such as reduced visibility in the readout of
$\ket{\downarrow\uparrow}$ or the initialization of
$\ket{\uparrow\downarrow}$.

\section{\label{conclusion}Conclusion}

We have developed a method that uses the measurement of a Rabi
resonance in the quantum--dot spin qubit to narrow the distribution of
the nuclear spin states. This method relies on Rabi oscillations
induced  via an oscillation of the singlet--triplet splitting $J$ in
the subspace $S^z=0$ of two electrons in a double quantum dot forming
a two--qubit system. Further, we have calculated several correlators
in the $S^z=0$ subspace for static $J$ and found that the transverse
components of pseudo--spin have a slower decay than the longitudinal
one. We have also discussed the implementation and fidelity of the
$\sqrt{\mathrm{SWAP}}$--gate in this system and the initialization to
the $\ket{\uparrow\downarrow}$, $\ket{\downarrow\uparrow}$ states.

\begin{acknowledgments}
We thank G. Burkard, M. Duckheim, J. Lehmann, F. H. L. Koppens,
D. Stepanenko and, in particular, A. Yacoby for useful
discussions. We acknowledge financial support from the Swiss NSF,
the NCCR nanoscience, EU NoE MAGMANet, DARPA, ARO, ONR, JST ICORP,
and NSERC of Canada.
\end{acknowledgments}

\appendix
\section{Drift in $\delta h^z$\label{App:chemicalshift}}
In addition to spin diffusion, driven by the nuclear dipole-dipole interaction, there may 
also be a change in $\delta h^z$ due to corrections to the projected effective Hamiltonian 
considered here (see Ref. \onlinecite{Coish:2005a}, Appendix B for details).  
After tracing out the electron pseudo--spin in state $\rho_S$, these correction 
terms give rise to an electron-mediated nuclear spin-spin interaction which, in general, takes 
the form of an anisotropic (XYZ) Heisenberg interaction
\begin{equation}
H_{nn} = \mathrm{Tr}_S\{\rho_S H \}= \sum_{i,j,\alpha=\{x,y,z\}} J^\alpha_{ij} I^\alpha_i I^\alpha_j.
\end{equation}
Here, the indices $i$ and $j$ run over all nuclear spin sites.

We use the corrections to leading order in the inverse Zeeman splitting $1/\epsilon_z$ 
($\epsilon_z=g\mu_B B$) given in Ref. \onlinecite{Coish:2005a}. This gives the 
typical value of the exchange constants $\left|J^\alpha_{ij}\right|\sim A^2/N^2\epsilon_z$.  
Assuming an unpolarized nuclear spin state, each nuclear spin will therefore precess in an 
effective mean field generated by all other spins in the dot of typical magnitude
\begin{equation}
h_{\mathrm{eff}} \sim \sqrt{N}\left|J^\alpha_{ij}\right| \sim A^2/N^{\frac{3}{2}}\epsilon_z.
\end{equation} 
This effective field will result in precession of the nuclear spins about an arbitrary angle 
(and hence, may change the value of $\delta h^z$) on a time scale
\begin{equation}\label{Eq:appendix}
\tau_p \sim N^{\frac{3}{2}}\epsilon_z/A^2 \sim 10^{-2}\,\mathrm{s},
\end{equation}
where we have assumed $N=10^6$ nuclear spins within the quantum dot, and 
$\epsilon_z/g\mu_B = A/g\mu_B \simeq 3.5\,\mathrm{T}$ for the time estimate. 
This is only a worst--case estimate, which neglects the effects, e.g., of a 
Knight-shift gradient (due to strong confinement of the electron), which may 
further weaken the dynamical effect discussed here.  We expect the dipolar nuclear 
spin diffusion time to be the limiting time scale for nuclear spin dynamics, in 
light of experiments on diffusion near donor impurities in GaAs.\cite{Paget:1982a} 
If the effect giving rise to $\tau_p$ in Eq. (\ref{Eq:appendix}) were significant, it could 
be further suppressed by choosing a larger quantum dot size or 
stronger magnetic field, thus allowing many electron spin measurements on the 
time scale of variation of $\delta h^z$. 

\section{Measurement\label{app:measurement}}
In this appendix we describe how a single measurement of the
two--electron system affects the nuclear spin state. We give the
analytical expression for the diagonal elements of the nuclear spin
density operator after a measurement. 

At t=0 the system is described by the following density operator
\begin{equation}
\rho(0)=\rho_e(0)\otimes\rho_I(0)= \ket{+}\bra{+}\otimes
\sum_ip_i\ket{\psi_I^i}\bra{\psi_I^i},
\end{equation}
with nuclear spin state $\ket{\psi_I^i}=\sum_n a_n^i\ket{n}$. The Hamiltonian $H_0$ of 
Eq.(\ref{Hamiltonian}) acts on the the nuclear--spin system as
$H_0\ket{n}=H_n\ket{n}$, where in $H_n$ the operator $\delta h^z$ has been replaced by $\delta
h^z_n$ (because $\delta h^z\ket{n}=\delta h^z_n\ket{n}$). Since $[H_0,\delta
h^z]=0$, only the diagonal elements of the nuclear density
operator $\rho_I$ (in the basis of $\delta h^z$) enter in matrix elements for
operators acting only on the two--electron system. As described in
Section \ref{sub:superposition}, these diagonal elements
$\rho_I(n)=\rho_I(n,0)=\bra{n}Tr_e\{\rho(0)\}\ket{n}$ 
describe a continuous Gaussian distribution in the continuum
limit. The trace over the electron system is 
defined as $Tr_e \rho(t)=\bra{+}\rho(t)\ket{+}
+\bra{-}\rho(t)\ket{-}$ and for $\rho_I(n,0)$ we have 
\begin{equation}
\rho_I(n,0)=\sum_i p_i |a_n^i|^2.
\end{equation}
The time evolution operators $U(t)$ and $U_n(t)$ are defined through $i\dot{U}(t)=H_0(t)U(t) $ and
$i\dot{U}_n(t)=H_n(t)U_n(t)$ and thus the density operator $\rho(0)$ evolves under the
Hamiltonian $H_0$ as
\begin{eqnarray}
\rho(t)&=&U(t)\rho(0)U^{\dagger}(t)\nonumber\\
&=&U(t)\left(\rho_e(0) \otimes\sum_i\sum_{n,l} p_i a_n^i
{a_l^i}^*\ket{n}\bra{l}\right)U^{\dagger}(t)\nonumber\\
&=&\sum_{n,l}\left(U_n(t)\rho_e(0)U^{\dagger}_l(t)\otimes\sum_ip_i
  a_n^i {a_l^i}^*\ket{n}\bra{l}\right).\nonumber \\ 
\end{eqnarray}
At time $t_m$ a measurement in the basis of $\ket{+}$
and $\ket{-}$ is performed on one single
two-electron system coupled to nuclear spins. Since
the outcome of this measurement is known, the state of the system after
the measurement is\cite{mixedmeasured} (the result depends on whether
$\ket{+}$ or $\ket{-}$  was measured) 
\begin{eqnarray}
\rho^{(1,\pm)}(t_m)&=&\frac{\ket{\pm}\bra{\pm}\rho(t_m)\ket{\pm}\bra{\pm}}
{P^{\pm}(t_m)}\nonumber \\
&=&\sum_{n,l}\left(\ket{\pm}\bra{\pm}
U_n(t_m)\rho_e(0)U^{\dagger}_l(t_m)\ket{\pm}\bra{\pm}\right.\nonumber\\
& &\,\,\,\,\,\,\,\,\,\,\,\,\,\,\otimes \sum_i\left.p_i
  a_n^i{a_l^i}^*\ket{n}\bra{l}\right)\frac{1}{P^{\pm}(t_m)},\nonumber \\ 
\end{eqnarray}
with
\begin{eqnarray}
P^{\pm}(t_m)&=&Tr_ITr_e\{\ket{\pm}\bra{\pm}\rho(t_m)\}\nonumber \\
&=&\sum_i\sum_n\frac{1}{2}\left(1\pm\langle\tau^z(t_m)\rangle_n\right)p_i|a_n^i|^2,\nonumber \\
\end{eqnarray}
where $Tr_I A=\sum_n\bra{n}A\ket{n}$ and
$\langle\tau^z(t)\rangle_n$ is given in Eq.(\ref{eq:tauz}). Here, $P^{\pm}(t_m)$ is the
probability to measure $\ket{\pm}$ at time $t_m$. We are mainly interested in
the diagonal elements of the nuclear density operator $\rho_I$ 
after the measurement. 
\begin{eqnarray}
\rho_{I}^{(1,\pm)}(n,t_m) &=& \bra{n}Tr_e\rho^{(1,\pm)}(t_m)\ket{n}\nonumber \\
 &=& \frac{\rho_I(n,0)}{P^{\pm}(t_m)}\bra{\pm}
 U_n(t_m)\rho_e(0)U^{\dagger}_n(t_m)\ket{\pm} \nonumber \\
 &=& \frac{\rho_I(n,0)}{
  P^{\pm}(t_m)}\frac{1}{2}\left(1 \pm \langle\tau^z(t_m)\rangle_n\right).
\end{eqnarray}
Using Eq.(\ref{eq:tauz}) we find
\begin{eqnarray}
\rho_{I}^{(1,+)}(n,t_m)&=&\frac{\rho_I(n,0)}{
  P^{+}(t_m)}\frac{1}{2}\left(\frac{2 (\Omega_n-\omega)^2} 
 {(\Omega_n-\omega)^2+(j/2)^2}\right.\nonumber\\
& &\left.+\frac{ (j/2)^2(1+\cos(\omega't_m))}{(\Omega_n-\omega)^2+(j/2)^2}\right)
\end{eqnarray}
and
\begin{equation}
\rho_{I}^{(1,-)}(n,t_m)=\frac{\rho_I(n,0)}{
  P^{-}(t_m)}\frac{1}{2}\frac{(j/2)^2(1-\cos(\omega't_m))} 
 {(\Omega_n-\omega)^2+(j/2)^2},
\end{equation}
where $\omega'$ is given in Eq.(\ref{eq:omegaprime}) and depends on the eigenvalue 
$\delta h^z_n$ of the nuclear spin eigenstate through $\Omega_n$. 

Parenthetically, we note that in the case (not described in this article) where
the measurement is performed on an ensemble of many 
different double quantum dots, the state of the ensemble after
the measurement is \cite{Peres:1993a}
\begin{eqnarray}
\rho^{(1)}_{ens}(t_m)&=&\sum_{n,l}\left(\ket{+}\bra{+}
U_n(t_m)\rho_e(0)U^{\dagger}_l(t_m)\ket{+}\bra{+}\right.\nonumber\\
& &+\left.\ket{-}\bra{-}
U_n(t_m)\rho_e(0)U^{\dagger}_l(t_m)\ket{+}\bra{+}\right)\nonumber\\
& &\otimes\sum_ip_i a_n^i{a_l^i}^*\ket{n}\bra{l},
\end{eqnarray}
and the nuclear--spin distribution has not changed. If a complete
measurement of the Rabi--resonance lineshape would be
performed on an ensemble of double dots, the result would be the Voigt
profile described in Sec. \ref{sub:superposition}.

\bibliography{ddhfxor}

\begin{thebibliography}{44}
\bibitem[{tim()}]{timeresolution}
\bibinfo{note}{{By "low time resolution", we mean that the measurement is
  performed at an unknown time $t_m$ (giving rise to state $\rho(t_m)$) in the
  interval $\Delta t=t_b-t_a$, with a uniform probability density $1/\Delta t$.
  The state after the measurement is then $\frac{1}{\Delta t}\int^{t_b}_{t_a}
  dt_m \rho(t_m)$.}}
\bibitem[{ass()}]{assumption}
\bibinfo{note}{{This assumption is not necessary for our narrowing scheme.
  However, it does allow for the derivation of the analytical formulas in this
  section, which give insight into the mechanism of narrowing. In the case of
  perfect time resolution, one would have an additional factor of $(1-\cos
  (\omega ' t_m))$ in Eq. (\ref{Eq:lorentzian}) ($\omega '$ is given in
  Eq.(\ref{eq:omegaprime})) and would also have to take into account the time
  $t_m$ at which each measurement was performed.}}
\bibitem[{typ()}]{typical}
\bibinfo{note}{{We have performed more than 60 runs of the simulation, varying
  $M$ and $j/\sigma_0$}}.
\bibitem[{mix()}]{mixedmeasured}
\bibinfo{note}{{The measurement performed at $t_m$ is not sufficient to
  distinguish the different states $\ket{\psi_I^i}$. This is because it is a
  measurement in the two--electron system and it only has an effect on the
  nuclear spin system in a way that some eigenstates $\ket{n}$ gain weight and
  some loose. This happens for all states $\ket{\psi_I^i}$ in the same way and
  thus from such a measurement we cannot tell into which of the states
  $\ket{\psi_I^i}$ the system has collapsed.}}
\end{thebibliography}

\end{document}